\documentclass[a4paper,11t]{article}
\usepackage[top=2.54cm,bottom=2.54cm, left=2.54cm, right=2.54cm]{geometry}
\usepackage{amsmath}
\usepackage{amsfonts}
\usepackage{amssymb}
\usepackage{graphicx}
\usepackage{bm}

\usepackage{enumerate}
\usepackage{color}
\usepackage[sort,comma]{natbib}
\usepackage{float}
\usepackage[colorlinks=true, allcolors=blue]{hyperref}
\usepackage{changepage}
\usepackage{booktabs}
\usepackage{relsize}

\makeatletter
\def\vec{\mathop{\operator@font vec}\nolimits}
\makeatother

\title{Constrained least squares simplicial-simplicial regression}
\author{Michail Tsagris \\
\\
Department of Economics, University of Crete, \\
Gallos Campus, Rethymnon, Greece  \\
\href{mailto:mtsagris@uoc.gr}{mtsagris@uoc.gr} 
}

\begin{document}

\maketitle

\begin{center}
{\bf Abstract}
\end{center}
Simplicial-simplicial regression refers to the regression setting where both the responses and predictor variables lie within the simplex space, i.e. they are compositional. For this setting, constrained least squares, where the regression coefficients themselves lie within the simplex, is proposed. The model is transformation-free but the adoption of a power transformation is straightforward, it can treat more than one compositional datasets as predictors and offers the possibility of weights among the simplicial predictors. Among the model's advantages are its ability to treat zeros in a natural way and a highly computationally efficient algorithm to estimate its coefficients. Resampling based hypothesis testing procedures are employed regarding inference, such as linear independence, and equality of the regression coefficients to some pre-specified values. The strategy behind the formulation of the new model is implemented is related to an existing methodology, that is of the same spirit, showcasing how other similar models can be employed as well. Finally, the performance of the proposed technique and its comparison to the existing methodology takes place using simulation studies and real data examples. 

\section{Introduction}
Compositional data\footnote{In the field of econometrics they are known as multivariate fractional data \citep{mullahy2015,murteira2016}.} are non-negative multivariate vectors whose variables (typically called components) conveying only relative information. When the vectors are scaled to sum to 1, their sample space is the standard simplex 
\begin{eqnarray} \label{simplex}
\mathbb{S}^{D-1}=\left\lbrace(y_1,...,y_D)^\top \bigg\vert y_i \geq 0,\sum_{i=1}^Dy_i=1\right\rbrace, 
\end{eqnarray}
where $D$ denotes the number of components. 

Examples of such data may be found in many different fields of study and the extensive scientific literature that has been published on the proper analysis of this type of data is indicative of its prevalence in real-life applications\footnote{For a substantial number of specific examples of applications involving compositional data see \cite{tsagris2020}.}. 

The widespread occurrence of this type of data in numerous scientific fields that involve predictors has necessitated the need for valid regression models which in turn has led to several developments in this area, many of which have been proposed recently. Most of these regression models have a restricted attention to the case of a simplicial response (simplicial-real regression setting), or a simplicial predictor (real-simplicial regression setting). The case of simplicial-simplicial regression, where both sides of the equation contain compositional data has not gained too much attention, and this is the main focus of this paper. 

Most published papers regarding the last case scenario involve transformations of both simplicial sides. \cite{hron2012}, \cite{wang2013}, \cite{chen2017} and \cite{han2022} used a log-ratio transformation for both the response and predictor variables and performed a multivariate linear regression model. \cite{alenazi2019} transformed the simplicial predictor using the $\alpha$-transformation \citep{tsagris2011} followed by principal component analysis and then employed the Kullback-Leibler divergence regression (or multinomial logit) model \citep{murteira2016}. The exception is \cite{fiksel2022} who proposed a transformation-free linear regression (TFLR) model whose coefficients lie within the simplex and are estimated via minimization of the Kullback-Leibler divergence (KLD) between the observed and the fitted simplicial responses. 

An important issue with compositional data analysis is the presence of zeros that prohibit the use of the logarithmic transformations, hence the approach of \cite{hron2012}, \cite{wang2013} and \cite{chen2017} cannot be applied, an issue that is not addressed in most papers. The classical strategy addressing this issue is to replace the zero values by a small quantity \citep{ait2003}. However, the approach of \cite{alenazi2019} handles the zero cases in a natural manner. This is not true in general for the TFLR model though.     

\cite{tsagris2011} categorized the compositional data analysis approaches into two main categories, the raw data approach and the log-ratio approach. A perhaps better classification would be the raw data and the transformation-based approaches. Moving along the lines of the raw data approach the paper proposes the use of the same transformation-free linear regression model, as in \cite{fiksel2022}, when both the response and the predictor variables are simplicial. However, the adoption of a power transformation in the simplicial response generalizes the model. The regression coefficients are estimated via simplicial constrained least squares (SCLS) and as the name implies, least squares is the loss function used to estimate the regression coefficients which are constrained to lie on the simplex. This in turn implies that the expected value of the simplicial response can be expressed as a Markov transition from the simplicial predictor. The proposed SCLS model allows for more than one simplicial predictor, further allows the possibility of assigning weights to the simplicial predictors, and treats zero values naturally, in both the simplicial response and the predictor variables. The assumption of linear independence between the simplicial variables, and hypotheses regarding the matrix of regression coefficients can be tested using resampling techniques. Evidently, the SCLS is similar in spirit to the TFLR, but they have different loss (or objective functions). Lastly, the TFLR model employs the Expectation-Maximization (EM) algorithm, whereas the SCLS model is based on quadratic programming, thus it enjoys a really low computational cost. 

The problem of constrained least squares (CLS), with a univariate real response, is not new. \cite{liew1976} and \cite{wets1991} have studied the asymptotic properties of constrained regression and have established the consistency of the regression coefficients, assuming the linear specification is correct. \cite{wets1991} specifically formalized the asymtptotic properties of the regression coefficients for the case of the M-estimators, whose least squares is a special case. More recently, \cite{james2019} proposed the constrained LASSO, a penalized version of the CLS. The current work though differs from these works in that it deals with the case of a constrained multivariate response.  

The rest of the paper is structured as follows. Section 2 reviews some simplicial-simplicial regression models, while Section 3 introduces the SCLS model, and discusses several cases related to the TFLR model as well. Section 4 contains Monte-Carlo simulation studies comparing the SCLS to the TFLR model in terms of a) type I and type II errors of the linear independence assumption, b) discrepancy of the estimated regression coefficients, and c) computational cost. The SCLS model is then applied to real data for illustration, and comparison to the TFLR model, purposes (Section 5), while the last section concludes the paper.  

\section{Review of simplicial-simplicial regression models}
Two commonly used log-ratio transformations, as well as a more general $\alpha$--transformation are defined, followed by some existing regression models for simplicial-simplicial regression.

\subsection{Log-ratio simplicial-simplicial regression models}
\cite{ait1982} suggested applying the additive log-ratio (alr) transformation to compositional data prior to using standard multivariate data analysis techniques. Let ${\bf y}=\left(y_1,\ldots, y_D\right)^\top \in \mathbb{S}^{D-1}$, then the alr transformation is given by 
\begin{eqnarray} \label{alr}
\mathbf{v} =\left\{v_j \right \}_{j=1,\ldots,D-1} = \left \{ \log \frac{y_j}{y_1} \right \}_{j=2,\ldots,D},
\end{eqnarray}
where ${\bf v}=\left(v_1,\ldots,v_{D-1}\right) \in \mathbb{R}^{D-1}$. Note that the common divisor, $y_1$, need not be the first component and was simply chosen for convenience. 

An alternative transformation proposed by \cite{ait1983} is the centred log-ratio (clr) transformation defined as 
\begin{eqnarray} \label{clr}
\mathbf{u} = \left \{u_j \right \}_{j=1,\ldots,D} = \left \{ \log{\frac{y_j}{g\left({\bf y}\right)}} \right \}_{j=1,\ldots,D,}
\end{eqnarray}
where $g\left({\bf y}\right)=\prod_{j=1}^Dy_j^{1/D}$ is the geometric mean. 

The clr transformation (\ref{clr}) was proposed in the context of principal component analysis with the potential drawback that $\sum_{j=1}^D u_j=0$, so essentially the unity sum constraint is replaced by the zero sum constraint. To overcome the singularity problem, \cite{ilr2003} proposed multiplying Eq. (\ref{clr}) by the $(D-1) \times D$ Helmert sub-matrix $\bf H$ \citep{helm1965,dryden1998,le1999}, an orthogonal matrix with the first row omitted, which results in what is called the isometric log-ratio (ilr) transformation
\begin{eqnarray} \label{ilr}
\mathbf{z}_{0} = {\bf H}\mathbf{u},
\end{eqnarray}
where $\mathbf{z}_{0} = \left(z_{0,1},\ldots, z_{0,D-1}\right)^\top \in \mathbb{R}^{D-1}$. Note that $\bf H$ may be replaced by any orthogonal matrix which preserves distances \citep{tsagris2011}. 

Simplicial-simplicial regression based on the ilr tranformation \citep{hron2012,wang2013,chen2017, han2022} is similar to alr regression and is carried out by transforming both the response and the predictor variables via the alr (\ref{alr}) or the ilr (\ref{ilr}) transformations
\begin{eqnarray*}
\text{E}({\bf v}_k|{\bf X})=\beta_{0k}+\sum_{j=1}^{D_p-1}\beta_{jk}\text{alr}({\bf X}_j) \ \ \text{and} \ \ \text{E}({\bf z}_{0k}|{\bf X})=\beta_{0k}+\sum_{j=1}^{D_p-1}\beta_{jk}\text{ilr}({\bf X}_j), \ \ k=1,\ldots,D_r-1, 
\end{eqnarray*}
where $D_r$ and $D_p$ denote the number of components of the response and compositional variables, respectively, and ${\bf v}_k$ and ${\bf z}_{0k}$ denote the $k$-variable of the alr (\ref{alr}) and ilr (\ref{ilr}) transformed simplicial response, respectively.

Moving along the same lines, \citep{wang2009} proposed partial least squares regression where both the response and the predictor variables are first transformed using the ilr transformation (\ref{ilr}), and the PLS is applied to the transformed data. Kernel regression \citep{dimarzio2015}, such as the Nadaraya-Watson or local polynomial regression, may also be applied in a similar manner. 

The fitted values for both the alr and ilr transformations are the same and are therefore generally back transformed onto the simplex using the appropriate inverse transformation for ease of interpretation. The drawback of these regression models is their inability to handle zero values in the compositional response data. The popular solution is to apply zero substitution strategies \citep{ait2003,martin2012} prior to fitting these regression models. 

\subsection{Kullback-Leibler divergence principal component regression}
Zero imputation strategies strategy, however, can produce regression models with predictive performance worse than regression models that can treat zeros naturally \citep{tsagris2015a}. When zeros occur in the data or more flexibility is required, the Box-Cox type transformation proposed by \cite{tsagris2011} may be employed. Specifically, \cite{ait2003} defined the power transformation as
\begin{eqnarray} \label{alpha}
{\bf w}_{\alpha}=\left\lbrace\frac{y_j^{\alpha}}{\sum_{l=1}^Dy_l^{\alpha}}\right\rbrace_{j=1,\ldots,D}
\end{eqnarray}
and \cite{tsagris2011} subsequently defined the $\alpha$-transformation, based on (\ref{alpha}), as
\begin{eqnarray} \label{isoalpha}
{\bf z}_{\alpha}=\frac{1}{\alpha}{\bf H}\left(D{\bf w}_{\alpha}-{\bf 1}_D\right), 
\end{eqnarray} 
where ${\bf H}$ is the Helmert sub-matrix and ${\bf 1}_D$ is the $D$-dimensional vector of 1s. 

While the power transformed vector ${\bf w}_{\alpha}$ in Eq. (\ref{alpha}) remains in the simplex $\mathbb{S}^{D-1}$, ${\bf z}_{\alpha}$ in Eq. (\ref{isoalpha}) is mapped onto a subset of $\mathbb{R}^{D-1}$. Furthermore, as $\alpha \rightarrow 0$, Eq. (\ref{isoalpha}) converges to the ilr transformation\footnote{The scaling factor $D$ exists to assist in the convergence. See \cite{tsagris2016} for more information.} in Eq. (\ref{ilr}) \citep{tsagris2016}, provided no zero values exist in the data. For convenience purposes, $\alpha$ is generally taken to be between $-1$ and $1$, but when zeros occur in the data, $\alpha$ must be restricted to be strictly positive. 

The benefit of the $\alpha$-transformation over the alr and clr transformations is that it can be applied even when zero values are present in the data (using strictly positive values of $\alpha$), offer more flexibility and yield better results \citep{tsagris2015b,tsagris2016,tsagris2022}.

In the context of simplicial-real regression, \cite{murteira2016} minimize the KLD
\begin{eqnarray} \label{klreg}
\min_{\pmb{\beta}} \sum_{i=1}^n{\bf y}_i^\top\log{\frac{{\bf y}_i}{\hat{\bf y}_i}}=
\max_{\pmb{\beta}} \sum_{i=1}^n{\bf y}_i^\top\log{\hat{{\bf y}}_i}, 
\end{eqnarray}
where $n$ denotes the sample size of the ${\bf u}_i$ the observed simplicial response data and $\hat{{\bf u}}_i=\left(\hat{y}_{i1}, \ldots, \hat{y}_{iD}\right)^\top$ are the fitted simplicial response data which have been transformed to simplex space through the transformation
\begin{eqnarray}  \label{regalpha}
\hat{y}_{ij}=
\left\lbrace
\begin{array}{cc} 
\frac{1}{1+\sum_{l=2}^{D_r}e^{{\bf x}_i^\top\pmb{\beta}_l}} & \text{if} \ \ j=1 \\
\frac{e^{{\bf x}_i^\top\pmb{\beta}_j}}{1+\sum_{l=2}^{D_r}e^{{\bf x}_i^\top\pmb{\beta}_l}} & \text{for} \ \ j=2,...,D_r,
\end{array} 
\right\rbrace
\end{eqnarray}
where where ${\bf x}^\top_i$ denotes the $i$-th row of the design matrix $\bf{X}$ containing the, non-simplicial, predictor variables, and $\pmb{\beta}_j=\left(\beta_{0j},\beta_{1j},...,\beta_{pj} \right)^\top$,  $j=2,\ldots,D$ \citep{tsagris2015a,tsagris2015b}.

The KLD regression model in Eq. (\ref{klreg}), also referred to as Multinomial logit regression is a semi-parametric regression technique, and unlike alr and ilr regression it can handle zeros naturally, since $\lim_{x \rightarrow 0}x\log{x}=0$.

\cite{alenazi2019} performed principal component analysis \citep{jolliffe2005} on the $\alpha$-transformed simplicial predictor $\bf X$, and then used the projections onto the first $k$ principal components as Euclidean predictors to the KLD regression model. This approach focuses on improving the predictive performance of the alr and ilr regression models, while increasing the computational cost, as the value of $\alpha$ and the number $k$ of the principal components must be tuned via cross-validation\footnote{For a comparison of this approach to the TFLR model, the reader is addressed to \cite{fiksel2022}.}.

\subsection{The TFLR model} \label{tflr}
The TFLR model \citep{fiksel2022} relates the $k$-th simplicial response vector to the simplicial predictor via a linear transformation 
\begin{eqnarray} \label{lt}
E({\bf Y}_k|{\bf X})=\sum_{j=1}^{D_p}B_{jk}{\bf X}_j,
\end{eqnarray}
where $\bf B$ itself belongs to the simplex, ${\bf B} \in \mathbb{R}^{D_p \times D_r} | B_{jk} \geq 0, \sum_{k=1}^{D_r}B_{jk}=1$. 
Estimation of the elements of $\bf B$ takes place by minimizing the KLD as in Eq. (\ref{klreg}) 
\begin{eqnarray} \label{kld}
\min\left\lbrace \sum_{i=1}^n\sum_{k=1}^{Dr}y_{ik}\log\left(\frac{y_{ik}}{\sum_{j=1}^{D_p}B_{jk}x_{ij}}\right)\right\rbrace = \max\left\lbrace \sum_{i=1}^n\sum_{k=1}^{Dr}y_{ik}\log\left(\sum_{j=1}^{D_p}B_{jk}x_{ij}\right) \right\rbrace.
\end{eqnarray}
The aforementioned approach allows for zero values in the response variable, but not for the simplicial predictor, in general. Think for example the case of an observation ${\bf x}_i=(x_{i1}, x_{i2}, 0, 0)$, where $x_{i1}$ and $x_{i2}$ are different from zero and and the $k$-th column of the estimated matrix $\bf B$ is equal to $(0, 0, B_{3k}, B_{4k})^\top$, where both $B_{3k}$ and $B_{4k}$ are not zero. Another possibility is when a column of the $\bf B$ matrix is full of zeros. These two cases will produce $x_{ik}B_{ik}=0$ and hence (\ref{kld}) cannot be computed\footnote{This phenomenon may occur with the KLD regression model as well but it is rather highly unlikely.}.  

\section{The simplicial constrained least squares model}
The SCLS model adopts the same link as TFLR (\ref{lt}) between the simplicial response and predictor variables, but only this time the elements of $\bf B$ are estimated by minimizing the squared loss
\begin{eqnarray} \label{cls}
\text{SL}({\bf B}) &=& \sum_{i=1}^n\sum_{k=1}^{D_r}\left(y_{ik}-\sum_{j=1}^{D_p}B_{jk}{\bf X}_j\right)^2
= \text{tr}\left({\bf Y} - {\bf X}{\bf B} \right)\left({\bf Y} - {\bf X}{\bf B} \right)^\top \nonumber \\
& \propto & 2\left[ - \text{tr}\left({\bf Y}^\top{\bf X}{\bf B}\right) + \frac{1}{2}\text{tr}\left({\bf B}^\top {\bf X}^\top {\bf X} {\bf B}\right)\right].
\end{eqnarray}

\subsection{The SCLS model and quadratic programming}
Quadratic programming solves the following problem
\begin{eqnarray} \label{qp}
\min_{\bm{b}} \left\lbrace-\bm{d}^\top \bm{b} + \frac{1}{2}\bm{b}^\top{\bf D}\bm{b} \right\rbrace, \ \ 
\text{under the constraints} \ \ {\bf A}^\top\bm{b} \geq \bm{b}_0.
\end{eqnarray}

The \text{SL}({\bf B}) though (\ref{cls}) is minimized via quadratic matrix programming, but it can be formulated under the quadratic programming framework\footnote{The SCLS model, along with all relevant functions used throughout this paper, has been implemented in the \textit{R} package \textit{Compositional} \citep{compositional2024} which makes use of the \textit{R} package \textit{quadprog} \citep{quadprog2019} that has implemented the algorithm of \cite{goldfarb1983} to minimize Eq. (\ref{cls}). Note that this implementation allows for equality constrains as well.}, as in the constrained minimization formula in Eq. (\ref{qp}). The $\bf D$ matrix is a $D_{rp} \times D_{rp}$ diagonal matrix, where $D_{rp}=D_r \times D_p$, and is related to the ${\bf X}^\top{\bf X}$ in the following manner
\begin{eqnarray} \label{Dmat}
{\bf D} = {\bf I}_{D_r} \otimes {\bf X}^\top{\bf X} =
\left( \begin{array}{cccc}
{\bf X}^\top{\bf X} &  {\bf 0} & \ldots & {\bf 0}\\
{\bf 0} & \ddots & {\bf 0} & \vdots \\
\vdots & {\bf 0} & \ddots &  {\bf 0} \\
{\bf 0} & \ldots & {\bf 0} &  {\bf X}^\top{\bf X} 
\end{array} \right),
\end{eqnarray}
where $\otimes$ denotes the Kronecker product and ${\bf I}_{D_r}$ is the $D_r \times D_r$ identity matrix. 

The ${\bf A}^\top$ matrix can be broken down two three sub-matrices, the $D_{rp} \times D_p$ ${\bf A}_1$, the $D_{rp} \times D_{rp}$ ${\bf A}_2$ and the the $D_{rp} \times D_{rp}$ ${\bf A}_3$, one stacked under the other, 
\begin{eqnarray*}
{\bf A}^\top=\left(\begin{array}{c} \Huge{\bf A}_1^\top \\ {\bf A}_2^\top \\ {\bf A}_3^\top \end{array} \right)
={\bf A}^\top = \left( \begin{array}{ccc}
{\bf I}_{D_p} & \ldots & {\bf I}_{D_p} \\
& {\mbox{\larger[2]$\textbf{I}_{D_{rp}}$}} & \\
& {\mbox{\larger[2]$-\textbf{I}_{D_{rp}}$}} &
\end{array}  \right).
\end{eqnarray*}
 
\begin{itemize}
\item The ${\bf A}_1^\top$ matrix contains the identity matrix ${\bf I}_{D_p}$, $D_r$ times, one next to the other
\begin{eqnarray*}
{\bf A}_1^\top = \left({\bf I}_{D_p}, \ldots, {\bf I}_{D_p}\right).    
\end{eqnarray*}
This is related to the unity sum constraint of the row coefficients. This means that
\begin{eqnarray*}
{\bf A}_1^\top {\bf b}=\left({\bf I}_{D_p}, \ldots, {\bf I}_{D_p}\right){\bf b}=\left(1,\ldots,1\right)^\top.
\end{eqnarray*}
\item The ${\bf A}_2^\top$ matrix is the identity matrix ${\bf I}_{D_rp}$
\begin{eqnarray*}
{\bf A}_2^\top = {\bf I}_{D_{rp}}.    
\end{eqnarray*}
This is related to the fact that all coefficients take non-negative values, thus
\begin{eqnarray*}
{\bf A}_2^\top {\bf b} = {\bf I}_{D_{rp}}{\bf b} \geq \left(0,\ldots,0\right)^\top.    
\end{eqnarray*}
\item The ${\bf A}_3^\top$ matrix is the negated identity matrix $-{\bf I}_{D_rp}$
\begin{eqnarray*}
{\bf A}_3^\top = -{\bf I}_{D_{rp}}.    
\end{eqnarray*}
\end{itemize}
This is related to the fact that the values of all coefficients must be less than or equal to 1, and hence
\begin{eqnarray*}
{\bf A}_3^\top {\bf b} = -{\bf I}_{D_{rp}}{\bf b} \geq \left(-1,\ldots,-1\right)^\top \ \ (\text{or} \ \ {\bf I}_{D_{rp}}{\bf b} \leq \left(1,\ldots,1\right)^\top).    
\end{eqnarray*}

The $\bm{b}$ relates to the $\bf B$ matrix via the vectorization operation $\bm{b}=\vec\left({\bf B}\right)$ and the $\bf d$ vector in Eq. (\ref{qp}) is the vectorized matrix ${\bf X}^\top{\bf Y}$, ${\bf d} = {\bf X}^\top{\bf Y}$. Finally, the $\bm{b}_0$ vector contains the $D_p$-dimensional vector of 1s, $\bm{j}_{D_p}$, that corresponds to the unity sum constraint of the rows of ${\bf B}$ (linked to the ${\bf A}_1^\top$ matrix), the $D_{rp}$-dimensional vector of 0s, $\bm{0}_{D_{rp}}$, corresponding to the fact that all elements $B_{jk}$ of ${\bf B}$ are non-negative (linked to the ${\bf A}_2^\top$ matrix) and  finally the $D_{rp}$-dimensional vector of -1s, $-\bm{j}_{D_{rp}}$, corresponding to the fact that all elements $B_{jk}$ of ${\bf B}$ are at bounded by unity from above (linked to the ${\bf A}_3^\top$ matrix)
\begin{eqnarray*} 
\bm{b}_0=\left(\bm{j}_{D_p}, \bm{0}_{D_{rp}}, -\bm{j}_{D_{rp}}\right)^\top.
\end{eqnarray*}

\subsection{Interpretation of the regression coefficients} \label{sec:interpretation}
Interpretation of the resulting estimated coefficients is a crucial aspect of a regression model if one is interested in making inference about them\footnote{Hypothesis testing for the parameters relies upon resampling techniques (presented next) due to the lack of parametric assumptions imposed on the coefficients. }. \cite{tsagris2023} for instance, proposed some non-parametric regression models that do not estimate coefficients, and thus visualized the (non-linear) effects of the predictor variables graphically. SCLS on the other hand yields regression coefficients and being a transformation free regression model, interpretation of its regression coefficients has similarities to the interpretation of the coefficients of the regression model with the alr (\ref{alr}) transformation. In the latter, a change in a predictor variable refers to a relative change in the components of the simplicial response, and in the SCLS, the interpretation is somewhat similar. If $x_j$ increases by $\delta$, while $x_l$ decreases by the same amount, ceteris paribus, the expected change of $\bf y$ is equal to $\delta\left({\bf B}_j -{\bf B}_l\right)$, where $j$ and $l$ denote rows. Assume for instance the following form of the matrix of regression coefficients
\begin{eqnarray} \label{Bmat}
{\bf B}=\left( \begin{array}{ccc}
0.20 & 0.40 & 0.40 \\
0.10 & 0.30 & 0.60 \\
0.30 & 0.35 & 0.35 \\
0.50 & 0.40 & 0.30 
\end{array}  \right)
\end{eqnarray}
If the first component of the predictor increases by $0.1$, while the second component decreases by the same amount, the expected change in the response is equal to $0.1 \left(0.2 - 0.1, 0.4 - 0.3, 0.4 - 0.6\right) = \left(0.01, 0.01, -0.02\right)$. While this interpretation is easy to understand it entails a perhaps minor downside. The interpretation is not universally applicable, as an expected change may lead to a point outside the simplex. 

\subsection{Visualization of the regression coefficients} \label{sec:visualization}
\begin{figure}[!ht]
\centering
\includegraphics[scale = 0.55, trim = 0 70 0 0]{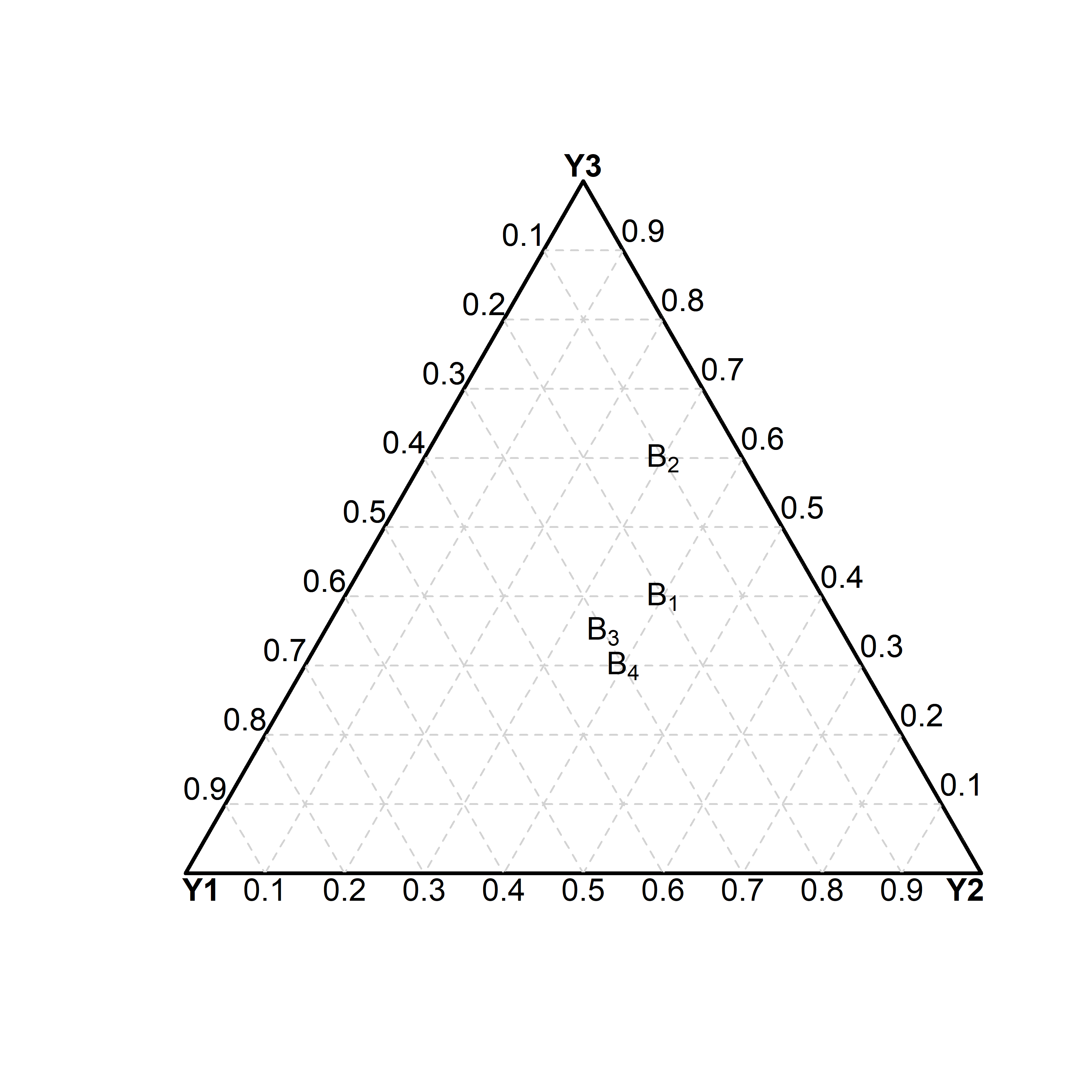} 
\caption{Ternary plot of the example matrix of coefficients $\bf B$ of Eq. (\ref{B}).}
\label{B}
\end{figure}

In the case of a 3-part simplicial response, the coefficients can also be visualized using the ternary diagram, as in Figure \ref{B}, that contains the 4 rows of the $\bf B$ matrix in Eq. (\ref{Bmat}). The interpretation of the ternary diagram is as follows. A point close to a vertex indicates high proportion in that vertex-associated component, while a point close to an edge opposite a vertex, indicates a low value in the vertex-associated component and finally a point that lies close to the barycentre of the triangle indicates almost equal values in all components. The third value of the second row of the matrix (${B}_{23}$) is equal to 0.60 and thus closer to the top vertex, whereas the third row values are similar, and hence B$_3$ lies close to the barycentre of the triangle. When the (row-)coefficients of a component of the simplicial predictor take values equal to the barycenter, i.e. $\left(1/D_r,\ldots,1/D_r\right)$ this indicates that the particular component of the simplicial predictor does not carry too much statistical information. On the contrary, the further the coefficients associated with a component of the simplicial predictor are away from the barycentre, the more information this component carries. In this simple example, the third component does not seem very important, because the third row coefficients (B$_3$) are close to the barycentre. The second component (B$_2$) of the simplicial predictor on the other hand, is probably the most important component, because its vector of coefficients lie the furthest from the barycentre among the other three vectors of coefficients.  

\subsubsection{Confidence intervals for the regression coefficients}
Confidence intervals for the true parameters can be constructed using non-parametric bootstrap \citep{fiksel2022} but the drawback is that the lower and upper values do not sum to unity. Confidence regions on the other hand are more intuitive. They are relatively easy to compute for 3-part simplicial responses and can be visualized in a ternary diagram in that case\footnote{In case of 4 components one could use a trinagular pyramid that has four equilateral triangles with all edges equal in length.}. Another non-parametric option would be to use empirical likelihood \citep{owen2001}, but to produce simultaneous confidence regions for all sets of regression coefficients would be computationally hard, especially as the dimensionality of the simplicial predictor, $D_p$, increases. To ease the computational burden one could produce profile confidence regions instead. Following \cite{fiksel2022}, the minimum volume ellipsoid \citep{van2009} that contains the 95\% of the bootstrap based coefficients is estimated and its ellipsoid hull is computed using the algorithm of \cite{pison1999}. 

\subsection{A test of linear independence}
Following \cite{fiksel2022} a test of linear independence between $\bf Y$ and $\bf X$ that relies upon permutations is proposed. If $\bf Y$ is linearly independent of $\bf X$, then $\text{E}({\bf Y}|{\bf X})=\text{E}({\bf Y})$. The test statistic utilized by \cite{fiksel2022} is based upon the KLD, but in the SCLS model it will be based upon the $\text{SL}^r\left({\bf B}\right)$. The steps to compute the permutation based p-value are delineated below.
\begin{enumerate}
\item[Step 1] Compute $\text{SL}^{obs}\left({\bf B}\right)$ using the observed data $\bf Y$ and $\bf X$.
\item[Step 2] Create ${\bf X}^r$ by permuting at random the rows of $\bf X$.
\item[Step 3] Compute $\text{SL}^r\left({\bf B}\right)$ using $\bf Y$ and ${\bf X}^r$.
\item[Step 4] Repeat Steps 2 and 3 $R$ times (e.g. $R=999$).
\item[Step 5] Calculate the permutation based $\text{p-value}=\frac{\sum_{r=1}^RI\left[ \text{SL}^r\left({\bf B}\right) \leq \text{SL}^{obs}\left({\bf B}\right) \right]+1}{R + 1}$.
\end{enumerate}

\subsection{A test for specific values imposed on regression coefficients}
In order to test the null hypothesis that ${\bf B}={\bf B}_0$ versus the alternative that at least one of the elements of $\bf B$ differs, the permutation technique may be employed as well. One such example is the case of $B_{j1}=\ldots=B_{jD_r}=1/D_r$ for some $j$, $1 \leq j \leq D_p$, indicating that the $j$-th component is not important. Below are two examples where this test could be applied. 

Amalagamation of components refers to summing the values of at least two components. For instance take the simplicial vector ${\bf x}=\left(x_1, x_2, x_3, x_4\right)$. An example of an amalgamation would be ${\bf x}^*=\left(x_1, x_2+x_3, x_4 \right)$. In the regression case this would imply that for some given components of the simplicial predictor, e.g. $1 \leq l_1, l_2 \leq D_p$ and $l_1 \neq l_2$, the expected value of the simplicial response is written as
\begin{eqnarray} \label{amalg}
E({\bf Y}_k|{\bf X})=\sum_{j \neq l_1, l_2}^{D_p}B_{jk}{\bf X}_j + B_{l_1k}{\bf X}_{l_1} + B_{l_2k}{\bf X}_{l_2} =\sum_{j \neq l_1, l_2}^{D_p}B_{jk}{\bf X}_j + B_{l_1k}\left({\bf X}_{l_1} +{\bf X}_{l_2}\right),
\end{eqnarray}
This implies to test for equality of two rows of the matrix $\bf B$, for instance that ${\bf B}_{l_1}={\bf B}_{l_2}$, where ${\bf B}_l$ denotes the $l$-th row of the matrix $\bf B$. The steps are similar to the linear test of independence where the difference is noticed at Step 2. Instead of permuting the rows of the simplicial predictor $\bf X$, its two columns, the components $\left({\bf x}_{l_1}, {\bf x}_{l_2}\right)$ are permuted. The rows of $\bf X$ are kept constant, but those two columns are permuted at random. 

If this hypothesis is not rejected it implies that these two components can be amalgamated and form a new simplicial predictor with $D_p-1$ components. Evidently, the same procedure could be applied to the simplicial response and in that case one ends up with an amalgamated simplicial response and in that case the hypothesis would refer to the equality of two columns of the matrix $\bf B$. 

A sub-composition stems from a composition by removing at least one component and rescaling the rest of the values to sum to unity. The resulting null hypothesis in this regression setting is $B_{j1}=\ldots=B_{jD_r}=0$ for some $j$, $1 \leq j \leq D_p$. However, there is an associated problem with this, the rescaling part, which in order to be valid the property of sub-compositional coherence\footnote{This property refers to the invariance of the results when the full composition or the sub-composition is used.} \citep{ait2003} should be met and in this case it is not. However, the theoretical implications of the sub-compositionality are not known, and hence this test perhaps requires more investigation. 

The hypothesis $B_{1k}=\ldots=B_{D_pk}=0$ for some $k$, $1 \leq k \leq D_r$ is also meaningful as it implies that the $k$-th component of the simplicial response is not affected by any component of the simplicial predictor. Note however that this hypothesis cannot be tested by the TFLR for the reason of introducing zeros to the fitted values. 

\subsection{More than one simplicial predictors}
The SCLS model can be extended to the case of $M$ ($M>1$) simplicial predictors, in which case Eq. (\ref{lt}) may be written as
\begin{eqnarray} \label{lt2}
E({\bf Y}_k|{\bf X}^1, \ldots, {\bf X}^M)=\sum_{m=1}^M\sum_{j=1}^{D_p}\frac{B_{jk}^m}{M}{\bf X}_j^m,
\end{eqnarray}
where the division of the matrix of coefficients $\bf B$ by $M$ ensures that the estimated simplicial responses sum to unity. 

\cite{fiksel2022} did not examine this case, but TFLR can be applied to this case as well. The drawback of the SCLS model in this case is that the $\bf D$ matrix (\ref{Dmat}) will not be positive definite, a drawback which can be handled by the algorithm of \cite{higham2002} (using Dykstra's correction) that forces positive definiteness. The algorithm yields an approximate solution which is deemed satisfactory. 

The implication of Eq. (\ref{lt2}) is that each simplicial predictor carries equal weight. One can escape this restrictive assumption by assigning weights to the regression coefficient matrix of each set of simplicial predictors, and hence write Eq. (\ref{lt2}) as
\begin{eqnarray} \label{lt3}
E({\bf Y}_k|{\bf X}^1, \ldots, {\bf X}^M)=\sum_{m=1}^M\sum_{j=1}^{D_p}a_m{B_{jk}^m}{\bf X}_j^m,   
\end{eqnarray}
where $a_m \geq 0$ for $m=1,\ldots,M$ and $\sum_{m=1}^Ma_m=1$, are the weights assigned to each set of simplicial predictors. Thus, the previous case (\ref{lt2}) can be seen as a special case of (\ref{lt3}) where $a_1=\ldots=a_m=1/M$. The weights may be computed using quadratic programming again, but in doing so the dimensionality of the required matrix $\bf D$ will explode as the number of simplicial predictors and the number of components of each increase, thus one can use simpler optimizers in \textit{R}. The possibility of allowing for more than one simplicial predictors in either model opens the door to conditional association testing and subsequently to simplicial variable selection. 

\subsection{The $\alpha$-SCLS and $\alpha$-TFLR models}
Another possible extension is to apply the power transformation (\ref{alpha}) to the simplicial response. 
\begin{eqnarray} \label{alt}
E({\bf W}_{\alpha}|{\bf X})={\bf XB},
\end{eqnarray}
and their predictions are then back-transformed using the inverse of (\ref{alpha})
\begin{eqnarray*} 
\hat{{\bf y}}=\left\lbrace\frac{\hat{w}_j^{1/\alpha}}{\sum_{l=1}^D\hat{w}_l^{1/\alpha}}\right\rbrace_{j=1,\ldots,D_r}.
\end{eqnarray*}
The extension will be denoted $\alpha$-SCLS and $\alpha$-TFLR models and evidently when $\alpha=1$ the SCLS and TFLR models, respectively, are recovered. The power transformation strategy was shown to improve the accuracy in the classification setting \citep{tsagris2014}, at the cost of introducing the interpretation-predictive performance trade-off. The cost of hard to interpret estimated regression parameters is compensated by the benefit of an increased predictive performance.  

\subsection{Examples of application of the SCLS model}
Similarly to the TFLR, the SCLS (and subesequently the $\alpha$-SCLS and $\alpha$-TFLR) can also be applied to a series of other regression settings. 

\subsubsection{AR(1) model}
\cite{fiksel2022} mentioned that in the AR(1) model formulation, the current expected simplicial response is a linear function of its lagged time response, $E({\bf Y}_t|{\bf Y}_{t-1})={\bf Y}_{t-1}{\bf B}$, thus the matrix $\bf B$ can be interpreted as the matrix of transition probabilities of the states of the simplicial response at time $t-1$. 
The linear independence test in this case translates to testing the assumption of no auto-correlation.
The case of allowing for more than one simplicial predictors extends the 1 lag to p lags, thus extend the AR(1) model to the AR(p) model. However, there is no theory to this end, what are implications of such a model (i.e. unit root testing, etc.) and thus this requires more research. 

\subsubsection{Categorical predictor or response}
Similarly to the TFLR, the case of a categorical predictor can be addressed via the SCLS. In that case, each row of the simplicial predictor is written as ${\bf x}_i={\bf e}_j$, where ${\bf e}_j$ denotes a vector with 0 $D_p$ elements except its $j$-th element that is 1\footnote{This could also be seen as an extreme case of compositional data, where one component takes a value of 1 and all other component have zero values.}, yielding a one-way analysis of variance type of model. 

The linear independence test translates to testing the assumption of equal simplicial population means. The latter case was studied by \cite{tsagris2017} using parametric and non-parametric procedures. However, simulation studies (not presented) replicating a case scenario from \cite{tsagris2017} comparing the SCLS and TFLR with the bootstrap version of the two sample James test showed that the latter is size correct, but the former two produced inflated type I errors. Hence, in contrast to the suggestion made by \cite{fiksel2022}, there is evidence against the use of SCLS and TFLR models for the purpose of hypothesis testing of equality of simplicial means. 

\cite{fiksel2022} stated that when the response is categorical this is equivalent to performing multinomial linear regression, with an identity link. The task is also equivalent to discriminant analysis for which better, non-linear, alternatives exist \citep{tsagris2016,lu2024}.

\subsection{Simplicial constrained median regression}
The $\bf D$ matrix in Eq. (\ref{Dmat}) is actually the cross-product of the following design matrix that contains the simplicial predictor
\begin{eqnarray} 
{\bf X}^* = {\bf I}_{D_r} \otimes {\bf X} =
\left( \begin{array}{cccc}
{\bf X} &  {\bf 0} & \ldots & {\bf 0}\\
{\bf 0} & \ddots & {\bf 0} & \vdots \\
\vdots & {\bf 0} & \ddots &  {\bf 0} \\
{\bf 0} & \ldots & {\bf 0} &  {\bf X} 
\end{array} \right).
\end{eqnarray}
The simplicial response can be vectorized ${\bf y}^* = \vec({\bf Y})$ and then estimate the coefficients of a logistic regression model of ${\bf y}^*$ on ${\bf X}^*$ with the identity as the link function, no constant term and imposing the same constraints as in the SCLS. This shows the equivalence between TFLR and univariate constrained logistic regression.

Further, this strategy is the same as the in one used in the seemingly unrelated regression equations (SURE) model \citep{zellner1962}. The SURE idea can also be used if one wants to substitute the square loss (\ref{cls}), or the KLD loss (\ref{kld}) with the $L_1$ norm loss and thus estimate the regression coefficients by minimizing the $L_1$ norm between the observed and the fitted simplicial responses, $\sum_{i=1}^n\sum_{j=1}^{D_r}|y_{ik}-\hat{y}_{ij}|$, and thus perform what may be termed as simplicial constrained median regression. The median regression has been implemented in the \textit{quantreg} \textit{R} package \citep{quantreg2023}. That implementation allows for inequality constraints and hence the constraints will be nearly the same to those used in Eq. (\ref{qp}). Evidently, all aforementioned generalizations (adoption of the $\alpha$-transformation and inclusion of multiple simplicial predictors) are applicable in this case as well.

\section{Simulation studies}
Simulation studies were conducted to compare the SCLS and TFLR regression models with the axes of comparison being a) the type I and II errors, b) the discrecpancy of the estimated matrix of coefficients, and c) the computational cost (running time). For the first two axes, the simplicial response consisted of $D_r = (3, 5, 7, 10)$ components, while the simplicial predictor contained only 3 components, and the sample sizes considered were $n=(50, 100, 200, 300, 500)$, but for the third axis, the sample size was larger. The results were averaged over $1,000$ replicates for each combination of dimensionality and sample size.

\subsection{Type I error}
Conforming to the null hypothesis of linear independence, both the values of the simplicial response $(\bf Y)$ and predictor $(\bf X)$ were generated from Dirichlet distributions, independently of one another. The values of the simplicial response were generated from a Dir$({\bf a})$, whose parameters ${\bf a}$ were drawn from the uniform distribution $U(1, 5)$, while the values of the simplicial predictor were generated from the Dir$(1, 1, 1)$. The estimated type I error for both models is presented in Table \ref{type1}, where evidently the results are similar, the permutation test attains the correct size, regardless of the model utilized, sample size and the dimensionality of the simplicial response.  

\begin{table}[ht]
\caption{Estimated type I error of both models, under the null hypothesis of linear independence.}
\label{type1}
\centering
\begin{tabular}{l|c|rrrr}
\toprule
            &       & \multicolumn{4}{c}{Number of response components} \\ \midrule
Sample size & Model &  $D_r=3$ & $D_r=5$ & $D_r=7$ & $D_r=10$ \\  \midrule
n=50  & SCLS & 0.046 & 0.050 & 0.055 & 0.045 \\ 
      & TFLR & 0.045 & 0.054 & 0.049 & 0.039 \\  \midrule
n=100 & SCLS & 0.038 & 0.058 & 0.050 & 0.040 \\ 
      & TFLR & 0.038 & 0.059 & 0.054 & 0.040 \\  \midrule
n=200 & SCLS & 0.058 & 0.047 & 0.057 & 0.050 \\ 
      & TFLR & 0.053 & 0.056 & 0.062 & 0.056 \\  \midrule
n=300 & SCLS & 0.046 & 0.043 & 0.050 & 0.044 \\ 
      & TFLR & 0.046 & 0.036 & 0.051 & 0.047 \\  \midrule
n=500 & SCLS & 0.046 & 0.046 & 0.052 & 0.051 \\ 
      & TFLR & 0.042 & 0.043 & 0.045 & 0.042 \\  \bottomrule
\end{tabular}
\end{table}

\subsection{Type II error}
The response now was linked to the predictor in a linear manner, as in Eq. (\ref{lt}) with some pre-determined values for $\bf B$ depending on $D_r$. Random vectors ${\bf x}_i$, for $i=1,\ldots,n$, were generated from $\text{Dir}\left(1,1,1\right)$, then transformed into $\pmb{\mu}_i={\bf x}_i{\bf B}$ and finally random vectors ${\bf y}_i$ were generated from $\text{Dir}\left(5\mu_1,\ldots,5\mu_{D_r}\right)$\footnote{For more information on the specific values of $\bf B$, the reader is referred to the Appendix.}. 

The estimated powers for both models are presented in Table \ref{type2}. For the small sample sized data, TFLR seems to produce higher estimated power levels than the SCLS, but as soon as the sample size increases their estimated power is equal 
\footnote{We also repeated the same experiment when the rows of the $\bf B$ matrix were nearly equal to $1/D_r$ but the power levels remained the same, indicating that the tests are really sensitive to small departures from the null hypothesis.}.

\begin{table}[ht]
\caption{Estimated power of both models.}
\label{type2}
\centering
\begin{tabular}{l|c|rrrr}
\toprule
            &       & \multicolumn{4}{c}{Number of response components} \\ \midrule
Sample size & Model &  $D_r=3$ & $D_r=5$ & $D_r=7$ & $D_r=10$ \\  \midrule
n=50  & SCLS & 0.977 & 1.000 & 0.996 & 0.942 \\ 
      & TFLR & 0.995 & 1.000 & 1.000 & 1.000 \\  \midrule
n=100 & SCLS & 1.000 & 1.000 & 1.000 & 1.000 \\ 
      & TFLR & 1.000 & 1.000 & 1.000 & 1.000 \\ \midrule
n=200 & SCLS & 1.000 & 1.000 & 1.000 & 1.000 \\ 
      & TFLR & 1.000 & 1.000 & 1.000 & 1.000 \\  \midrule
n=300 & SCLS & 1.000 & 1.000 & 1.000 & 1.000 \\ 
      & TFLR & 1.000 & 1.000 & 1.000 & 1.000 \\  \midrule
n=500 & SCLS & 1.000 & 1.000 & 1.000 & 1.000 \\ 
      & TFLR & 1.000 & 1.000 & 1.000 & 1.000 \\  \bottomrule
\end{tabular}
\end{table}

\subsection{Discrepancy of the estimated matrix of coefficients}
The discrepancy of the estimated matrix of coefficients was assessed by the KLD of the estimated to the true values of the $\bf B$ matrix, and the $L_1$ distance between the estimated and the true values of the $\bf B$ matrix 
\begin{eqnarray*}
\text{KLD}(\hat{\bf B}, {\bf B})=\sum_{j=1}^{D_p}\sum_{k=1}^{D_r}\hat{B}_{jk}\log\left(\frac{\hat{B}_{jk}}{B_{jk}}\right) \ \ \text{and}  \ \ L_1(\hat{\bf B}, {\bf B})=\sum_{j=1}^{D_p}\sum_{k=1}^{D_r}|\hat{B}_{jk}-B_{jk}|. 
\end{eqnarray*}

Table \ref{bias} contains the estimated discrepancy quantities. Both discrepancy metrics exhibit small differences between the two models, but the TFLR produces better results. However, the differences in the discrepancies are rather small and can be asserted that the two models are not substantially different. Secondly, for a given dimensionality of the simplicial response, as the sample size increases the differences, using both measures, between the two competing models diminish. Lastly, the KLD increases as the dimensionality of the simplicial responses increases, for a given sample size, however, the $L_1$ metric decreases, exhibiting a rather unexpected behaviour. 

\begin{table}[ht]
\caption{Estimated discrepancy of the regression coefficients of the SCLS and TFLR models.}
\label{bias}
\centering
\begin{tabular}{l|r|rrrr|rrrr}
\toprule
&       & \multicolumn{4}{c}{KLD$(\hat{\bf B}, {\bf B})$} & \multicolumn{4}{c}{$L_1(\hat{\bf B}, {\bf B})$} \\      
\midrule
& Model & $D_r=3$ & $D_r=5$ & $D_r=7$ & $D_r=10$ & $D_r=3$ & $D_r=5$ & $D_r=7$ & $D_r=10$ \\ 
\midrule
n=50  & SCLS & 0.0096 & 0.0208 & 0.0166 & 0.0163 & 0.0533 & 0.0379 & 0.0414 & 0.0307 \\ 
      & RFLR & 0.0087 & 0.0145 & 0.0143 & 0.0130 & 0.0514 & 0.0348 & 0.0400 & 0.0290 \\  \hline
n=100 & SCLS & 0.0048 & 0.0122 & 0.0088 & 0.0087 & 0.0380 & 0.0271 & 0.0299 & 0.0218 \\ 
      & TFLR & 0.0044 & 0.0085 & 0.0077 & 0.0070 & 0.0365 & 0.0249 & 0.0287 & 0.0206 \\  \hline
n=200 & SCLS & 0.0024 & 0.0075 & 0.0046 & 0.0050 & 0.0274 & 0.0201 & 0.0211 & 0.0158 \\ 
      & TFLR & 0.0022 & 0.0051 & 0.0040 & 0.0039 & 0.0263 & 0.0181 & 0.0202 & 0.0148 \\  \hline
n=300 & SCLS & 0.0014 & 0.0054 & 0.0031 & 0.0036 & 0.0215 & 0.0166 & 0.0172 & 0.0131 \\ 
      & TFLR & 0.0013 & 0.0037 & 0.0027 & 0.0029 & 0.0207 & 0.0149 & 0.0164 & 0.0123 \\  \hline
n=500 & SCLS & 0.0008 & 0.0037 & 0.0021 & 0.0024 & 0.0170 & 0.0132 & 0.0136 & 0.0103 \\ 
      & TFLR & 0.0007 & 0.0025 & 0.0018 & 0.0019 & 0.0162 & 0.0118 & 0.0129 & 0.0095 \\ 
\bottomrule
\end{tabular}
\end{table}

\subsection{Computational cost}
To evaluate the computational cost the command \textit{benchmark()}, from the \textit{R} package \textit{Rfast2} \citep{rfast2023}, was used, based on 20 repetitions. Simplicial response values with $D_r=(3, 5, 7, 10)$ and simplicial predictor values with $D_p=3$ were generated from a Dirichlet distribution and large sample sizes were considered, $n=(500, 1000, 5000, 10000, 20000)$. The speedup factors of the SCLS compared to TFLR are presented in Table \ref{durat}. The ratios clearly depict that the time required to fit the TFLR model is significantly higher compared to that of the SCLS model. 

The TFLR model has been implemented in the \textit{R} package \textit{codalm} \citep{codalm2021} and the code is written in \textit{R}, wheras the SCLS model uses the \textit{R} package \textit{quadprog} \citep{quadprog2019} that relies upon Fortran. So, the running time comparison is not completely fair. Secondly, the code in the TFLR implementation is not highly optimized. To address the second issue, the comparison took place using a self implementation of the TFLR model. In this implementation, the estimated regression coefficients from the SCLS model were used as starting values in the EM algorithm. The new speed-up factors of the SCLS compared to TFLR also appear in Table \ref{durat}. For the low dimensionality of the simplicial response, the self implementation can be up to 10 times faster than the \textit{codalm}'s implementation, whereas for a higher dimensionality, the improvement drops to only 4 times. 

A \textit{C++} implementation is expected to make the TFLR algorithm even faster. However, the number of computations involved in the EM algorithm is still high compared to the quadratic programming approach. Apart from the computational burden, the TFLR model may break down for some combination of zeros in the simplicial predictor and the matrix of coefficients $\bf B$ as mentioned in Section \ref{tflr}, whereas the SCLS model treats those cases naturally. 

\begin{table}[ht]
\centering
\caption{Speedup factors of SCLS compared to TFLR: ratio of running time of the TFLR model to the running time of the SCLS model.}
\label{durat}
\begin{tabular}{l|rrrr|rrrr}
\toprule
  & \multicolumn{8}{c}{Number of response components} \\ \midrule
  & \multicolumn{4}{c}{\textit{codalm} implementation} & \multicolumn{4}{c}{Self implementation of EM algorithm} \\ \midrule
Sample size & $D_r=3$ & $D_r=5$ & $D_r=7$ & $D_r=10$ & $D_r=3$ & $D_r=5$ & $D_r=7$ & $D_r=10$ \\  \midrule
n=500   & 52.430  & 41.701  & 58.552  & 68.708  & 11.216 & 15.406 & 20.166 & 28.943 \\ 
n=1000  & 69.645  & 85.490  & 67.583  & 94.403  & 12.156 & 21.820 & 27.049 & 66.684 \\ 
n=5000  & 121.677 & 114.959 & 160.351 & 261.529 & 12.151 & 51.776 & 37.861 & 78.321 \\ 
n=10000 & 201.123 & 222.886 & 159.677 & 231.467 & 24.744 & 37.547 & 59.314 & 83.515 \\ 
n=20000 & 196.560 & 329.418 & 207.237 & 213.228 & 26.963 & 41.129 & 42.529 & 56.916 \\  \bottomrule
\end{tabular}
\end{table}

\section{Real data analysis} 
Two real data sets are used to illustrate the performance of the SCLS and compare it with that of TFLR. The first data set comes from the field of agricultural economics while the second comes from the field of political sciences. A third data set was then used to illustrate the confidence regions (of the SCLS model) of the row coefficients of the matrix $\bf B$. Lastly, two of these datasets were further utilized to illustrate the performance of the $\alpha$-SCLS.

\subsection{Crop cultivated area and crop production}
Data regarding crop productivity in the Greek NUTS II region of Thessaly during the 2017-218 cropping year were supplied by the Greek Ministry of Agriculture, also known as farm accountancy data network (FADN) data. The data refer to a sample of 487 farms and initially they consisted of 20 crops, but after aggregation they were narrowed down to 10 crops\footnote{A larger version of this dataset was used in \citep{mattas2024}.}. For each of the 487 farms the cultivated area and the production in each of the 10 crops is known. However, the goal of the paper is to relate the composition of the production (simplicial response, $\bf Y$) to the composition of the cultivated area (simplicial predictor, $\bf X$) and for this reason were scaled to sum to unity\footnote{The raw data cannot be distributed due to disclosure restrictions.}. Table \ref{fadn} contains information about the data: the component names and their simple arithmetic averages.   

\begin{table}[ht]
\centering
\caption{Information on the FADN dataset.}
\label{fadn}
\begin{tabular}{l|cc}
\toprule
Component           & Cultivated area ($\bf X$) & Production ($\bf Y$) \\  \midrule
X1: Other Cereals   & 0.1063 & 0.0958 \\
X2: Durum Wheat     & 0.1539 & 0.1407 \\
X3: Maize           & 0.0600 & 0.1015 \\
X4: Potatoes, Protein Crops and Rice & 0.0393 & 0.0230 \\
X5: Cotton & 0.2315 & 0.2087 \\
X6: Tobacco, Oil Seeds, Industrial Crops and Vegetables & 0.0449 & 0.0708 \\
X7: Green Plants, Pasture and Grazing & 0.1777 & 0.1777 \\
X8: Fruits, Berries and Nuts & 0.0945 & 0.1098 \\
X9: Olive Trees & 0.0733 & 0.0446 \\
X10: Grapes and Wine & 0.0186 & 0.0273 \\ \bottomrule
\end{tabular}
\end{table}

Tables \ref{be1} contains the regression coefficients estimated via the SCLS and TFLR models. Evidently, the diagonal coefficients take high values as expected, as the proportion of each crop's cultivated area is expected to be highly related to the proportion of the same crop production. As expected, the linear independence hypothesis is rejected, based on both models. The assumption of a diagonal matrix of coefficients ${\bf B}={\bf I}_{10}$ was also rejected by both models. 

\begin{table}[ht]
\centering
\caption{FADN data: estimated regression coefficients of the SCLS and TFLR models.}
\label{be1}
\begin{tabular}{rrrrrrrrrrr}
\toprule
   & \multicolumn{10}{c}{Estimated coefficients based on the SCLS model} \\ \midrule
   & Y1 & Y2 & Y3 & Y4 & Y5 & Y6 & Y7 & Y8 & Y9 & Y10 \\ \midrule
X1 & 0.9246 & 0.0000 & 0.0342 & 0.0095 & 0.0000 & 0.0051 & 0.0078 & 0.0120 & 0.0000 & 0.0069 \\ 
X2 & 0.0000 & 0.9337 & 0.0310 & 0.0000 & 0.0000 & 0.0112 & 0.0163 & 0.0000 & 0.0000 & 0.0078 \\ 
X3 & 0.0000 & 0.0000 & 1.0000 & 0.0000 & 0.0000 & 0.0000 & 0.0000 & 0.0000 & 0.0000 & 0.0000 \\ 
X4 & 0.0210 & 0.0285 & 0.0063 & 0.5571 & 0.0000 & 0.1653 & 0.2217 & 0.0000 & 0.0000 & 0.0000 \\ 
X5 & 0.0000 & 0.0000 & 0.0427 & 0.0000 & 0.9229 & 0.0333 & 0.0011 & 0.0000 & 0.0000 & 0.0000 \\ 
X6 & 0.0000 & 0.0000 & 0.0000 & 0.0000 & 0.0000 & 1.0000 & 0.0000 & 0.0000 & 0.0000 & 0.0000 \\ 
X7 & 0.0000 & 0.0000 & 0.0649 & 0.0000 & 0.0000 & 0.0000 & 0.9351 & 0.0000 & 0.0000 & 0.0000 \\ 
X8 & 0.0000 & 0.0000 & 0.0000 & 0.0000 & 0.0000 & 0.0000 & 0.0000 & 1.0000 & 0.0000 & 0.0000 \\ 
X9 & 0.0232 & 0.0197 & 0.0068 & 0.0025 & 0.0117 & 0.0313 & 0.0115 & 0.1062 & 0.7485 & 0.0385 \\ 
X10 & 0.0000 & 0.0000 & 0.0000 & 0.0000 & 0.0000 & 0.0000 & 0.0000 & 0.0000 & 0.0000 & 1.0000 \\ 
\midrule
   & \multicolumn{10}{c}{Estimated coefficients based on the TFLR model} \\ \midrule
   & Y1 & Y2 & Y3 & Y4 & Y5 & Y6 & Y7 & Y8 & Y9 & Y10 \\ \midrule
X1 & 1.0000 & 0.0000 & 0.0000 & 0.0000 & 0.0000 & 0.0000 & 0.0000 & 0.0000 & 0.0000 & 0.0000 \\ 
X2 & 0.0000 & 0.9995 & 0.0000 & 0.0000 & 0.0000 & 0.0000 & 0.0000 & 0.0000 & 0.0000 & 0.0005 \\ 
X3 & 0.0000 & 0.0000 & 1.0000 & 0.0000 & 0.0000 & 0.0000 & 0.0000 & 0.0000 & 0.0000 & 0.0000 \\ 
X4 & 0.0000 & 0.0000 & 0.0000 & 0.8653 & 0.0000 & 0.0556 & 0.0791 & 0.0000 & 0.0000 & 0.0000 \\ 
X5 & 0.0000 & 0.0000 & 0.0000 & 0.0000 & 1.0000 & 0.0000 & 0.0000 & 0.0000 & 0.0000 & 0.0000 \\ 
X6 & 0.0000 & 0.0000 & 0.0000 & 0.0000 & 0.0000 & 0.9985 & 0.0000 & 0.0000 & 0.0015 & 0.0000 \\ 
X7 & 0.0000 & 0.0000 & 0.0000 & 0.0000 & 0.0000 & 0.0000 & 1.0000 & 0.0000 & 0.0000 & 0.0000 \\ 
X8 & 0.0000 & 0.0000 & 0.0000 & 0.0000 & 0.0000 & 0.0000 & 0.0000 & 0.9993 & 0.0007 & 0.0000 \\ 
X9 & 0.0000 & 0.0000 & 0.0000 & 0.0000 & 0.0000 & 0.0000 & 0.0000 & 0.0392 & 0.9608 & 0.0000 \\ 
X10 & 0.0000 & 0.0000 & 0.0000 & 0.0000 & 0.0000 & 0.0000 & 0.0000 & 0.0000 & 0.0000 & 1.0000 \\ 
\bottomrule
\end{tabular}
\end{table}

\begin{figure}[!ht]
\centering
\begin{tabular}{cc}
\includegraphics[scale = 0.55, trim = 0 0 0 0]{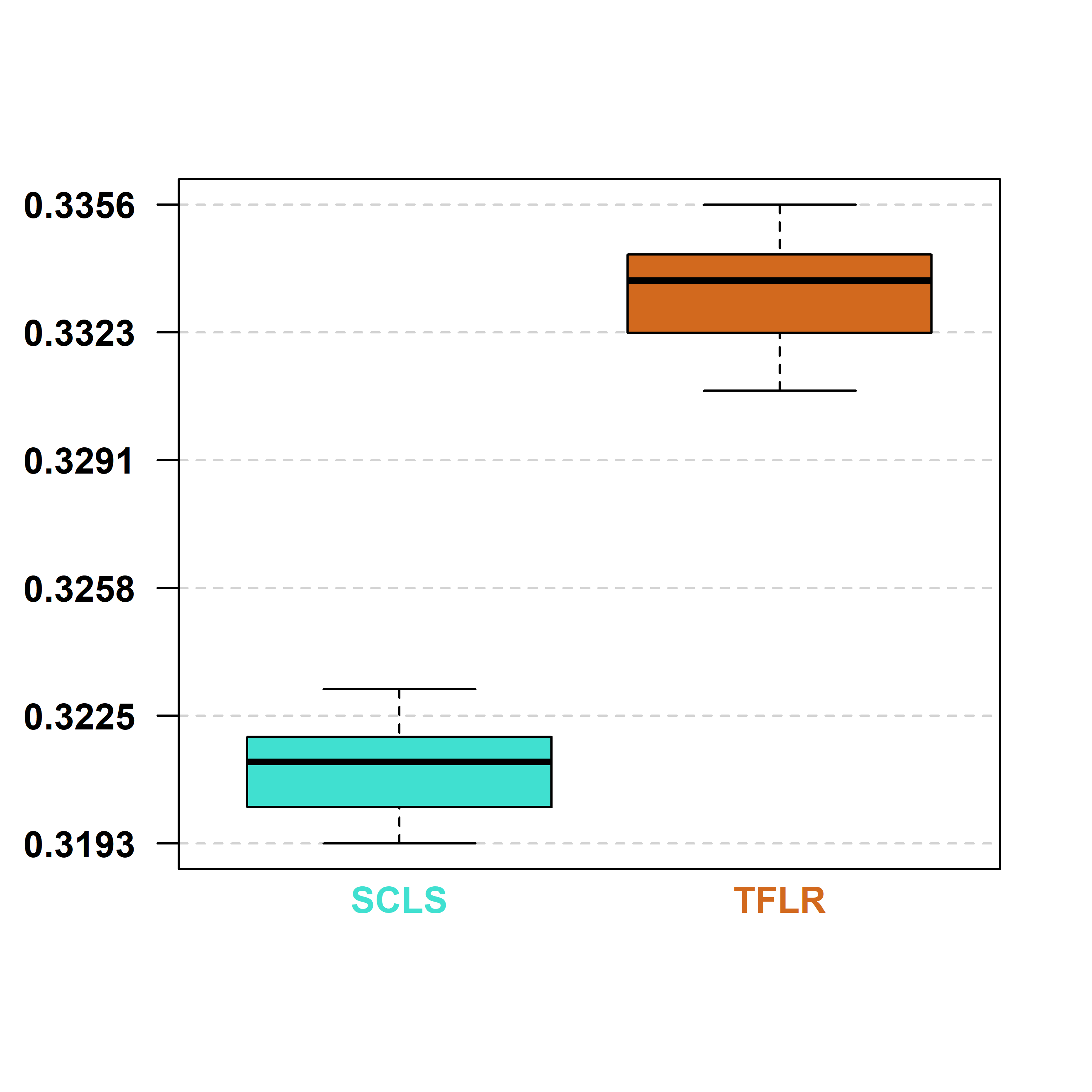}  &
\includegraphics[scale = 0.55, trim = 40 0 0 0]{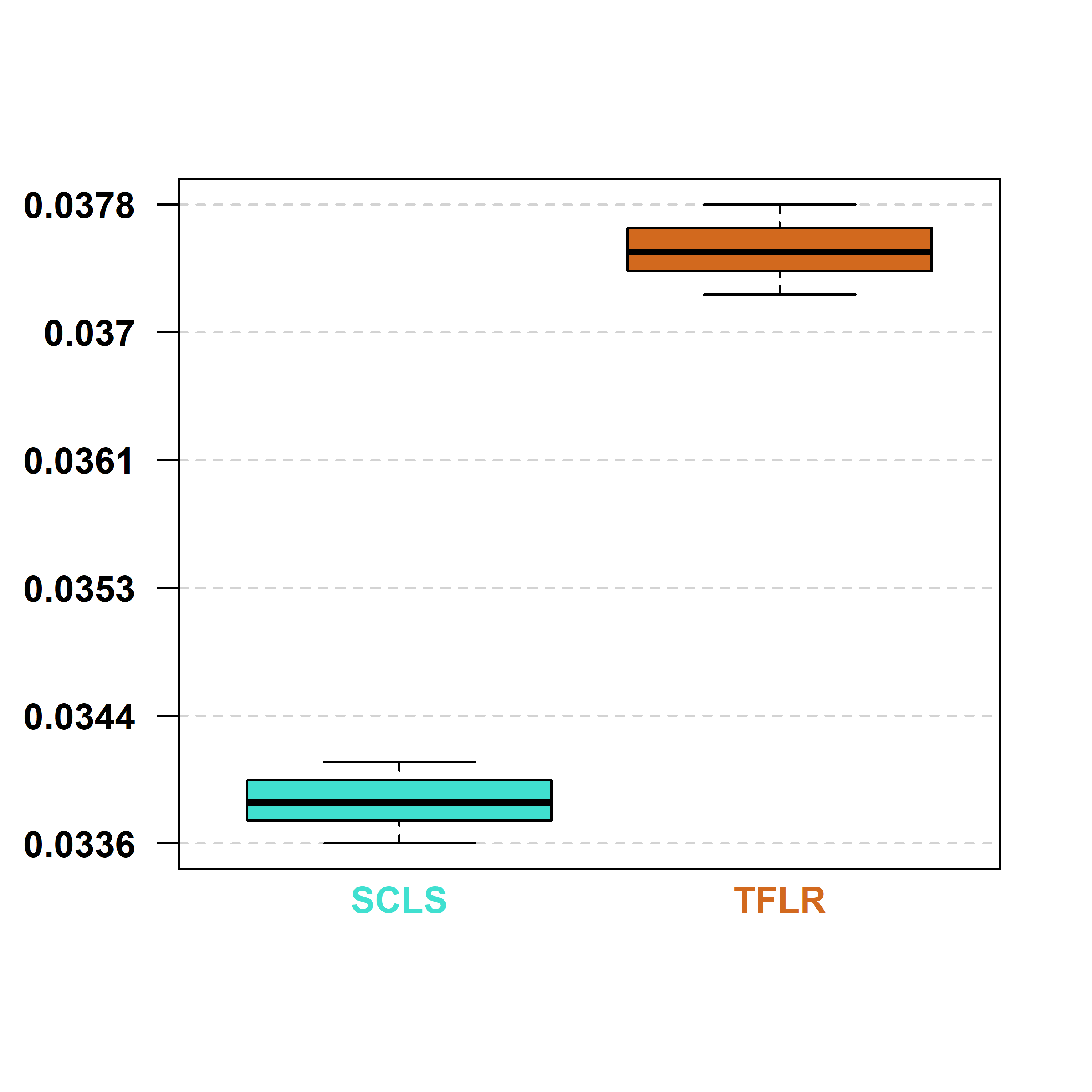}   \\
(a) KLD & (b) JSD 
\end{tabular}
\caption{FADN data: box-plots of the predictive KLD and JSD for the SCLS and TFLR models.}
\label{perf1}
\end{figure}

In order to assess the fit of the two models the KLD of the observed to the predicted simplicial response values and the Jensen-Shannon divergence (JSD) were employed. The KLD was 0.0267 for both models, whereas the JSD values were 0.025 and 0.028 for the SCLS and the TFLR models, respectively. To estimate the predictive performance of each model the 10-fold cross-validation (CV) was utilised and the performance metrics were again the KLD and the JSD. The 10-fold CV was repeated 20 times and the box-plots appearing in Figure \ref{perf1} visualize the results. The average KLD values were 0.321 and 0.33 for the SCLS and the TFLR models, respectively, whereas the average JSD values were 0.034 and 0.038 for the SCLS and the TFLR models, respectively. 

\subsection{Catalan elections}
The data set contains the votes in Catalan elections from year 1980 up to 2006 for 41 regions each year. The main parties consist of 6 candidates, while there are votes for other candidates, blank votes and null votes. In total there are 328 observations with 9 variables, that were scaled to sum to 1. The goal here is to assume an AR(1) model, where the lag refers to the one time period between two consecutive election years. Table \ref{elect} contains the averages in the 9 components classified per year of election.  

\begin{table}[ht]
\centering
\caption{Catalan elections data: yearly average voting proportions by candidate, blank and null votes.}
\label{elect}
\begin{tabular}{l|rrrrrrrrr}
\toprule
     & \multicolumn{9}{c}{Voting proportions} \\ \midrule
Year & CiU & PSC & PP & IC & ERC & CC & Other & Blank & Null \\ \midrule
1980 & 0.3095 & 0.1947 & 0.0157 & 0.1186 & 0.1018 & 0.0000 & 0.2482 & 0.0058 & 0.0058 \\ 
1984 & 0.5640 & 0.2208 & 0.0745 & 0.0370 & 0.0533 & 0.0000 & 0.0395 & 0.0047 & 0.0062 \\ 
1988 & 0.5385 & 0.2358 & 0.0503 & 0.0464 & 0.0573 & 0.0000 & 0.0588 & 0.0068 & 0.0060 \\ 
1992 & 0.5260 & 0.2250 & 0.0540 & 0.0397 & 0.1011 & 0.0000 & 0.0363 & 0.0116 & 0.0064 \\ 
1995 & 0.4869 & 0.2172 & 0.1022 & 0.0516 & 0.1227 & 0.0000 & 0.0064 & 0.0091 & 0.0038 \\ 
1999 & 0.4669 & 0.3025 & 0.0734 & 0.0076 & 0.1170 & 0.0000 & 0.0202 & 0.0089 & 0.0035 \\ 
2003 & 0.3877 & 0.2467 & 0.0875 & 0.0497 & 0.2039 & 0.0000 & 0.0116 & 0.0090 & 0.0039 \\ 
2006 & 0.3769 & 0.2329 & 0.0792 & 0.0720 & 0.1797 & 0.0125 & 0.0194 & 0.0209 & 0.0065 \\ 
\bottomrule
\end{tabular}
\end{table}

Since the task of interest is to perform a time series analysis, the AR(1) SCLS and TFLR models were fitted to the data from the years 1980 up to 2003, and the data from the year of elections 2006 was considered to be the test set. Table \ref{be2} contains the estimated coefficients of the SCLS and TFLR models. The prediction capabilities showed that in terms of the KLD, the SCLS was worse than the TFLR, 0.132 versus 0.080, respectively, whereas in terms of the JSD, the two models performed equally well, both produced JSD value equal to 0.002. 

\begin{table}[ht]
\centering
\caption{Catalan elections data: estimated regression coefficients of the SCLS and TFLR models.}
\label{be2}
\begin{tabular}{rrrrrrrrrr}
\toprule
              & \multicolumn{9}{c}{Estimated coefficients based on the SCLS model} \\ \midrule
              & CiU & PSC & PP & IC & ERC & CC & Other & Blank & Null \\  \midrule
CiU$_{t-1}$   & 0.8326 & 0.0392 & 0.0000 & 0.0000 & 0.0632 & 0.0000 & 0.0390 & 0.0165 & 0.0094 \\ 
PSC$_{t-1}$   & 0.0000 & 0.7031 & 0.0910 & 0.1502 & 0.0481 & 0.0000 & 0.0077 & 0.0000 & 0.0000 \\ 
PP$_{t-1}$    & 0.0000 & 0.4776 & 0.5000 & 0.0000 & 0.0224 & 0.0000 & 0.0000 & 0.0000 & 0.0000 \\ 
IC$_{t-1}$    & 0.1725 & 0.5814 & 0.0207 & 0.0738 & 0.0000 & 0.0000 & 0.1516 & 0.0000 & 0.0000 \\ 
ERC$_{t-1}$   & 0.1854 & 0.0000 & 0.0679 & 0.0000 & 0.7468 & 0.0000 & 0.0000 & 0.0000 & 0.0000 \\ 
CC$_{t-1}$    & 0.1111 & 0.1111 & 0.1111 & 0.1111 & 0.1111 & 0.1111 & 0.1111 & 0.1111 & 0.1111 \\ 
Other$_{t-1}$ & 0.9222 & 0.0000 & 0.0778 & 0.0000 & 0.0000 & 0.0000 & 0.0000 & 0.0000 & 0.0000 \\ 
Blank$_{t-1}$ & 0.0000 & 0.0000 & 1.0000 & 0.0000 & 0.0000 & 0.0000 & 0.0000 & 0.0000 & 0.0000 \\ 
Null$_{t-1}$  & 1.0000 & 0.0000 & 0.0000 & 0.0000 & 0.0000 & 0.0000 & 0.0000 & 0.0000 & 0.0000 \\ 
\midrule
              & \multicolumn{9}{c}{Estimated coefficients based on the TFLR model} \\ \midrule
              & CiU & PSC & PP & IC & ERC & CC & Other & Blank & Null \\ \midrule
CiU$_{t-1}$   & 0.8605 & 0.0357 & 0.0000 & 0.0000 & 0.0548 & 0.0000 & 0.0334 & 0.0101 & 0.0055 \\ 
PSC$_{t-1}$   & 0.0000 & 0.7452 & 0.0711 & 0.1368 & 0.0380 & 0.0000 & 0.0000 & 0.0088 & 0.0000 \\ 
PP$_{t-1}$    & 0.0000 & 0.4217 & 0.5342 & 0.0000 & 0.0272 & 0.0000 & 0.0000 & 0.0168 & 0.0000 \\ 
IC$_{t-1}$    & 0.1908 & 0.5020 & 0.0000 & 0.1158 & 0.0000 & 0.0000 & 0.1902 & 0.0000 & 0.0012 \\ 
ERC$_{t-1}$   & 0.1499 & 0.0000 & 0.0636 & 0.0000 & 0.7863 & 0.0000 & 0.0000 & 0.0002 & 0.0000 \\ 
CC$_{t-1}$    & 0.1111 & 0.1111 & 0.1111 & 0.1111 & 0.1111 & 0.1111 & 0.1111 & 0.1111 & 0.1111 \\ 
Other$_{t-1}$ & 0.8211 & 0.0000 & 0.1422 & 0.0000 & 0.0000 & 0.0000 & 0.0277 & 0.0000 & 0.0089 \\ 
Blank$_{t-1}$ & 0.0000 & 0.0000 & 1.0000 & 0.0000 & 0.0000 & 0.0000 & 0.0000 & 0.0000 & 0.0000 \\ 
Null$_{t-1}$  & 0.1904 & 0.0000 & 0.0000 & 0.2025 & 0.0000 & 0.0000 & 0.2198 & 0.0803 & 0.3071 \\
\bottomrule
\end{tabular}
\end{table}

\subsection{Education level of father and mother}
This data set contains the education level of father and mother in percentages of low (l), medium (m), and high (h) of 31 countries in Europe\footnote{The dataset is available from the \textit{R} package \textit{robCompositions} \citep{robCompositions2023}.} and the simplicial response is the father's education level, while the simplicial predictor is the mother's educational level. This data set was used by \cite{fiksel2022} and was chosen on the grounds of illustrating a) the interpretation of the coefficients (see Section \ref{sec:interpretation}), and b) their 95\% confidence regions (see \ref{sec:visualization}). 

\begin{figure}[!ht]
\centering
\includegraphics[scale = 0.55, trim = 0 70 0 0]{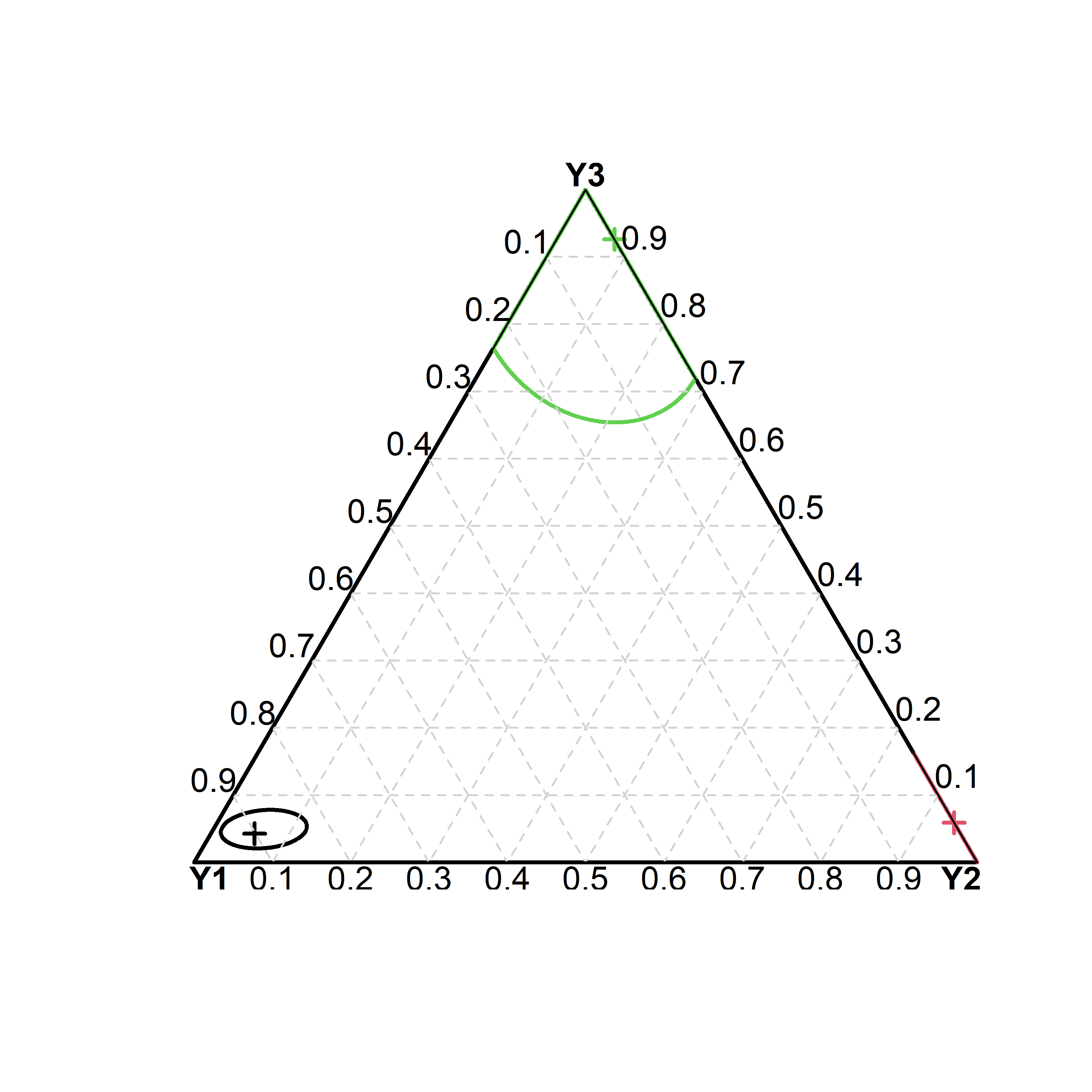} 
\caption{Ecucation level data: confidence regions of the coefficients of the SCLS model: the black refers to the first row coefficients, the red refers to the second row and the green refers to the third row of coefficients.}
\label{confr}
\end{figure}

The matrix of the estimated regression coefficients of the SCLS model and of the TFLR model (for comparison purposes) are given below
\begin{eqnarray*}
\hat{\bf B}_{SCLS}=\left( \begin{array}{ccc}
0.9014 & 0.0559 & 0.0428 \\
0.0000 & 0.9409 & 0.0591 \\
0.0000 & 0.0737 & 0.9263 
\end{array}  \right) \ \ \text{and}  \ \
\hat{\bf B}_{TFLR}=\left( \begin{array}{ccc}
0.9113 & 0.0512 & 0.0375 \\
0.0000 & 0.9054 & 0.0946 \\
0.0000 & 0.1415 & 0.8585
\end{array}  \right),
\end{eqnarray*}
where the rows correspond to the low, medium and high educational levels of the mother, whereas the columns indicate the same educational levels for the father. If the percentage of low educated mothers increases (additively) by $\delta$ while the percentage of medium educated mothers decreases (additively) by $\delta$, the expected change in the three educational levels of the father is $\delta \left(0.9014 - 0, 0.0559 - 0.9409, 0.0428 - 0.0591\right)=\left(0.914\delta, -0.885\delta, -0.0163\delta\right)$. Figure \ref{confr} presents the 95\% confidence regions of the three row coefficients. Evidently, there is high uncertainty in the coefficients corresponding to the highly educated women (third row coefficients) as depicted by the figure. 

\subsection{Predictive performance of the power transformed SCLS model}
\begin{figure}[!ht]
\centering
\begin{tabular}{cc}
\includegraphics[scale = 0.55]{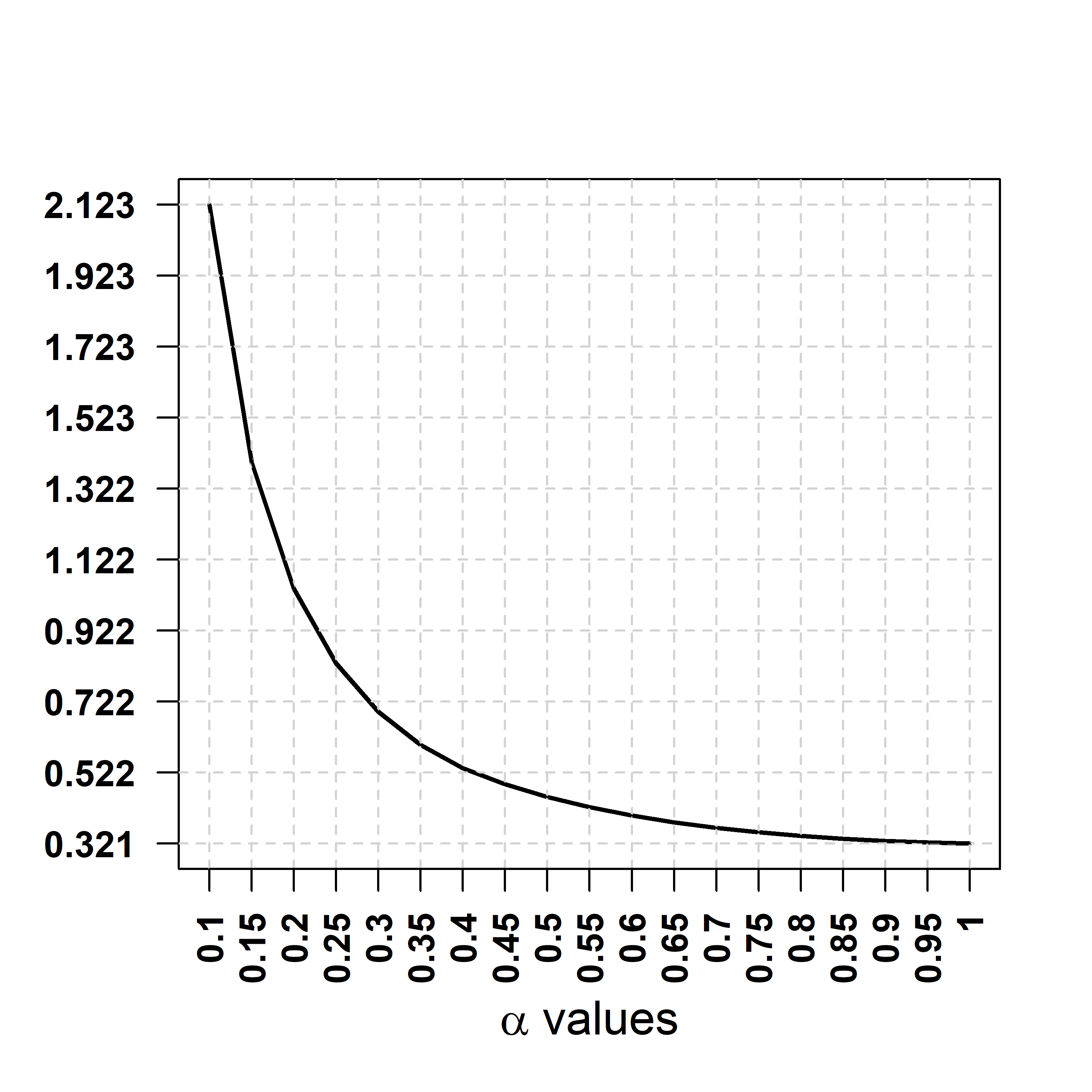}  &
\includegraphics[scale = 0.55]{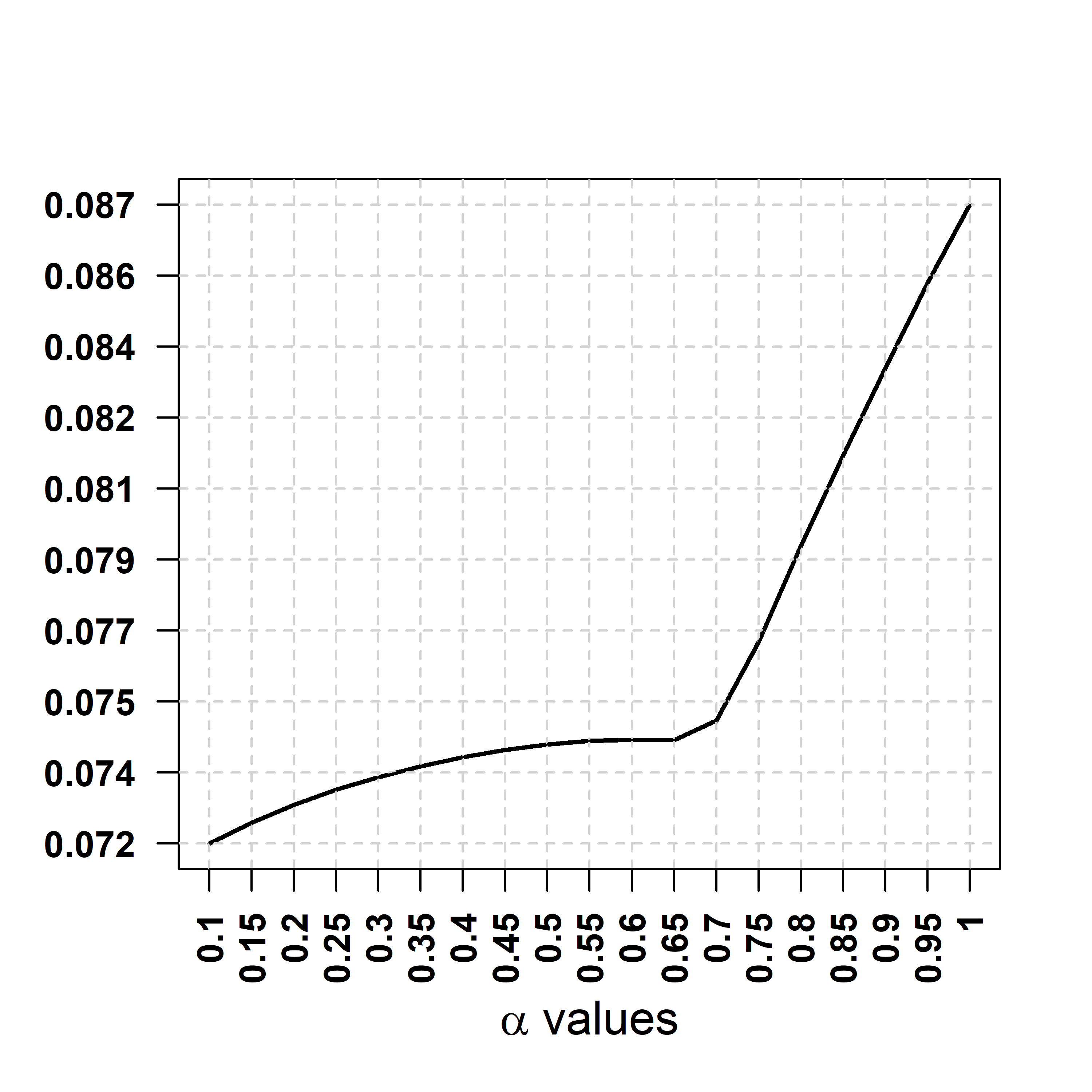}   \\
(a) FADN data & (b) Education data 
\end{tabular}
\caption{Predictive performance measured by KLD for the $\alpha$-SCLS model as a function of the $\alpha$-values.}
\label{perf2}
\end{figure}

The FADN and the education data sets were further used to test the performance of the SCLS model when applied to the power transformed data using Eq. (\ref{alpha}) employing the 10-fold CV protocol repeated 20 times. Figure \ref{perf2} presents the average predictive KLD measures as a function of the $\alpha$-values. Evidently, this strategy was not proved prosperous for the FADN data, but it was beneficiary in the case of the Education data set. 

\section{Conclusions}
The paper proposed the constrained least squares approach to estimate the parameters of a linear transformation-free regression model for compositional data with compositional predictors. This regression model was first proposed by \cite{fiksel2022} who, in contrast to this approach, estimated its parameters using the KLD from the observed to the fitted compositional data. This paper showed that extensions to multiple simplicial predictors can straightforwardly be adopted by either approach and further, the categorical predictors case was examined via small simulation studies and the evidence was against the use of this model. The extensive simulation studies comparing both approaches provided evidence that both models perform nearly equally well and there is no clear winner. Finally, the inclusion of a power parameter in the simplicial response provided evidence of increased predictive performance, at the cost of parameter interpretability. Thus, both approaches share similar properties and can be seen as two sides of the same coin. An advantage however of the SCLS over its competitor, the TFLR, is the first's high computational efficiency. 

Closing this paper we would like to pose some possible research directions and suggestions. We live in the era of big data and algorithms ought to be customizable to treat such cases. The SCLS implementation in the \textit{R} package \textit{Compositional} addresses this issue by accepting Filebacked Big Matrices. Another option would be to "cut" the data in chunks and via parallel computing perform the SCLS model in each chunk and then combine the results in a meta-analytic way. An alternative, and promising, strategy would be to perform subdata selection \citep{chasiotis2024}.

The SURE idea may be used for other multivariate regression models or loss functions as well. For instance, extending the $L_1$ and $L_2$ norm regression models to the general $L_p$ regression model \citep{money1982}. Other options include the substitution of the the KL divergence by the Jensen-Shannon divergence or any other $\phi$-divergence. The common ground in these aforementioned extensions is the set of constraints imposed on the matrix of regression coefficients $\bf B$ and the linear link between the response and predictor variables.

Finally some more suggestions are the following. (a) Investigation of any possible benefit of the power transformation (\ref{alpha}) to the simplicial predictor(s). This could yield more more accurate predictions, but at the cost of increased computational cost.  b) Exploitation of the relationship between compositional and directional data. By taking the square root both variables transform into directional data for which spherical regression models exist and in order to ensure that the fitted values will lie within the simplex the folded Kent model \citep{scealy2014} may be employed. c) Investigation of the ensemble learning in the $\alpha$-SCLS model. Instead of selecting one value of $\alpha$ combine the fitted values of many models produced by different values of $\alpha$. d) The possibility to add non-linear effects in the model, regardless of the loss function used, to increase the flexibility of the model and escape the assumption of linear relationship. For instance, the use of neural networks, or local polynomial regression with simplicial constraints imposed on the coefficients.

\clearpage
\section*{Appendix}
\setcounter{section}{0}
\renewcommand{\thesubsection}{A\arabic{subsection}}
\setcounter{equation}{0}
\renewcommand{\theequation}{\thesubsection.\arabic{equation}}
\setcounter{figure}{0}
\renewcommand\thefigure{A\arabic{figure}}    

\subsection{Example of the quadratic programming formulation}
Table \ref{qp} illustrates an example of the sub-matrices ${\bf A}_1^\top$, ${\bf A}_2^\top$ and ${\bf A}_3^\top$ of the matrix ${\bf A}^\top$ and the vector ${\bf b}_0$. In this case both simplicial response and predictor variables contain 3 components ($D_r=D_p=3$). 

\begin{eqnarray*}
{\bf A}^\top = 
\left( \begin{array}{c}
{\bf A}_1^\top \\
{\bf A}_2^\top \\
{\bf A}_3^\top \\
\end{array}
\right) = 
\left( \begin{array}{ccccccccc}
1 & 0 & 0 & 1 & 0 & 0 & 1 & 0 & 0  \\ 
0 & 1 & 0 & 0 & 1 & 0 & 0 & 1 & 0  \\ 
0 & 0 & 1 & 0 & 0 & 1 & 0 & 0 & 1  \\ \hline
1 & 0 & 0 & 0 & 0 & 0 & 0 & 0 & 0  \\ 
0 & 1 & 0 & 0 & 0 & 0 & 0 & 0 & 0  \\ 
0 & 0 & 1 & 0 & 0 & 0 & 0 & 0 & 0  \\ 
0 & 0 & 0 & 1 & 0 & 0 & 0 & 0 & 0  \\ 
0 & 0 & 0 & 0 & 1 & 0 & 0 & 0 & 0  \\ 
0 & 0 & 0 & 0 & 0 & 1 & 0 & 0 & 0  \\ 
0 & 0 & 0 & 0 & 0 & 0 & 1 & 0 & 0  \\ 
0 & 0 & 0 & 0 & 0 & 0 & 0 & 1 & 0  \\ 
0 & 0 & 0 & 0 & 0 & 0 & 0 & 0 & 1  \\ \hline
-1 & 0 & 0 & 0 & 0 & 0 & 0 & 0 & 0 \\ 
0 & -1 & 0 & 0 & 0 & 0 & 0 & 0 & 0 \\ 
0 & 0 & -1 & 0 & 0 & 0 & 0 & 0 & 0 \\ 
0 & 0 & 0 & -1 & 0 & 0 & 0 & 0 & 0 \\ 
0 & 0 & 0 & 0 & -1 & 0 & 0 & 0 & 0 \\ 
0 & 0 & 0 & 0 & 0 & -1 & 0 & 0 & 0 \\ 
0 & 0 & 0 & 0 & 0 & 0 & -1 & 0 & 0 \\ 
0 & 0 & 0 & 0 & 0 & 0 & 0 & -1 & 0 \\ 
0 & 0 & 0 & 0 & 0 & 0 & 0 & 0 & -1 \\ 
\end{array}  \right)^\top
\ \ \text{and} \ \ 
{\bf b}_0 =
\left( \begin{array}{c}
1 \\ 
1 \\
1 \\
0 \\
0 \\
0 \\
0 \\
0 \\
0 \\
0 \\
0 \\
0 \\
-1 \\
-1 \\
-1 \\
-1 \\
-1 \\
-1 \\
-1 \\
-1 \\
-1 \\
\end{array} \right) 
\end{eqnarray*}

\subsection{Information (entropy) of the coefficients}
\begin{figure}[!ht]
\centering
\includegraphics[scale = 0.8]{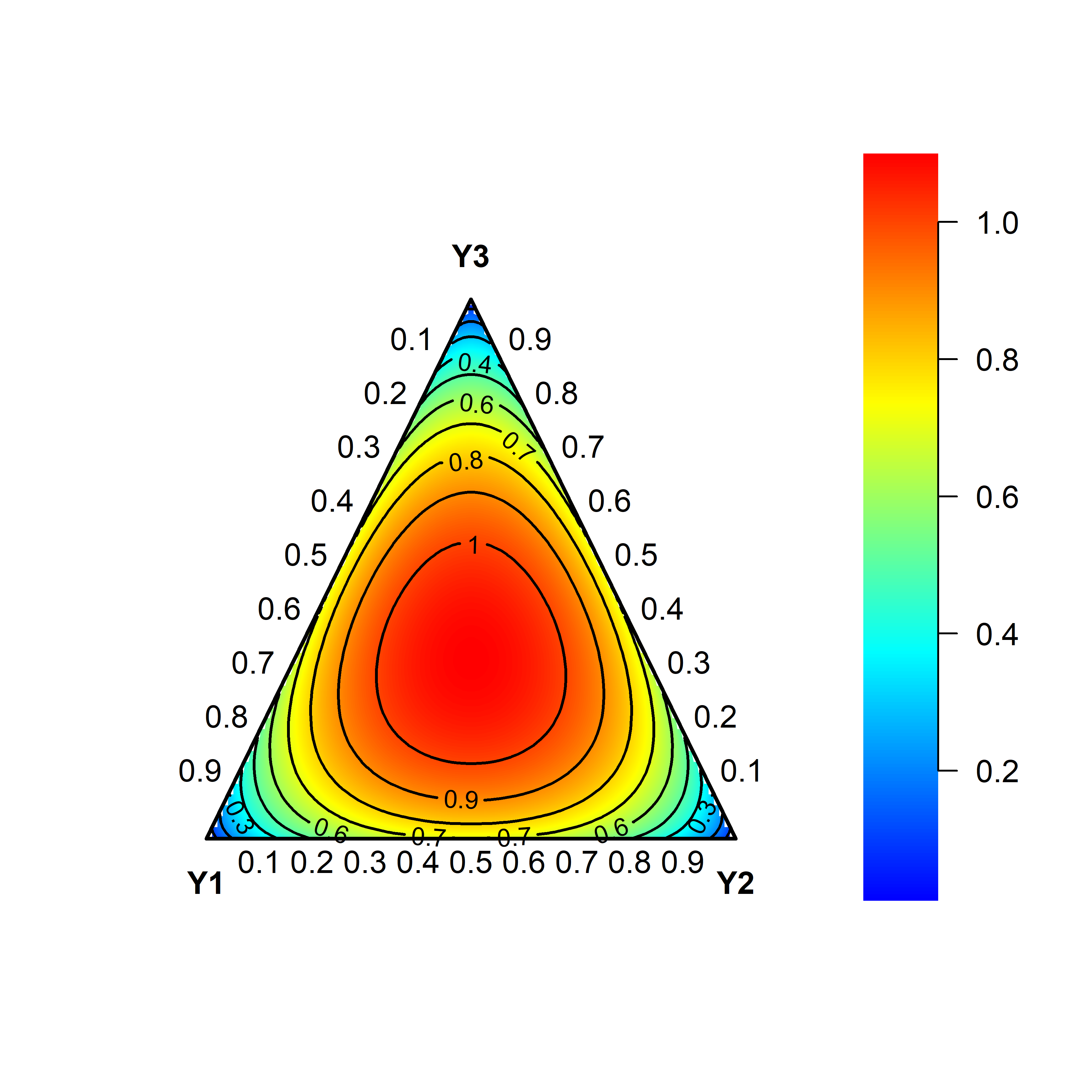} 
\caption{Negated entropy of data points in the simplex.}
\label{entropy}
\end{figure}

Figure \ref{entropy} shows the negated entropy of data points in the 3-part simplex, $\mathbb{S}^2$. The entropy is maximized (or the negated entropy is maximised) as we move towards the barycentre of the triangle, i.e. the center of the simplex. 

\subsection{A computationally efficient implementation of the linear test of independence}
Instead of calling the function \textit{scls()} from the \textit{R} package \textit{Compositional} multiple times, a function was created minimizing the necessity to perform computations. The vector ${\bf b}_0$ is always the same, and so is the matrix $\bf D$ since even by permuting the rows of the simplicial predictor $\bf X$ the cross-product ${\bf X}^\top{\bf X}$ remains the same. The only thing that changes is the $\bf d$ vector. Secondly, Eq. (\ref{cls}) was computed excluding the trace of the matrix ${\bf Y}{\bf Y}^\top$ as this is constant. 

\subsection{Details on the data generation for type II error and discrepancy of the estimated coefficients}
Random vectors ${\bf x}_i$, for $i=1,\ldots,n$, were generated from $\text{Dir}\left(1,1,1\right)$, then transformed into $\pmb{\mu}_i={\bf x}_i{\bf B}$ and finally random vectors ${\bf y}_i$ were generated from $\text{Dir}\left(5\mu_1,\ldots,5\mu_{D_r}\right)$. The following 4 matrices $\bf B$ were used. 

\begin{eqnarray*}
{\bf B}_3 &=& \left( \begin{array}{ccc}    
0.45 & 0.00 & 0.55 \\ 
0.20 & 0.34 & 0.46 \\ 
0.76 & 0.01 & 0.23 \\ 
\end{array} \right) \\ 
{\bf B}_5 &=& \left( \begin{array}{ccccc}    
0.31 & 0.00 & 0.04 & 0.65 & 0.01 \\ 
0.02 & 0.01 & 0.00 & 0.48 & 0.48 \\ 
0.28 & 0.02 & 0.64 & 0.06 & 0.00 \\ 
\end{array} \right) \\
{\bf B}_7 &=& \left( \begin{array}{ccccccc}    
0.16 & 0.20 & 0.00 & 0.11 & 0.32 & 0.12 & 0.09 \\ 
0.63 & 0.08 & 0.00 & 0.09 & 0.10 & 0.08 & 0.01 \\ 
0.10 & 0.24 & 0.20 & 0.12 & 0.03 & 0.01 & 0.30 \\ 
\end{array} \right) \\
{\bf B}_{10} &=& \left( \begin{array}{cccccccccc}    
0.25 & 0.00 & 0.01 & 0.09 & 0.01 & 0.00 & 0.24 & 0.14 & 0.00 & 0.26 \\ 
0.44 & 0.10 & 0.18 & 0.02 & 0.01 & 0.00 & 0.09 & 0.07 & 0.00 & 0.10 \\ 
0.34 & 0.03 & 0.00 & 0.14 & 0.17 & 0.00 & 0.04 & 0.00 & 0.19 & 0.09 \\ 
\end{array} \right) 
\end{eqnarray*}

\clearpage
\bibliographystyle{chicago}
\bibliography{vivlio}

\end{document}